\documentstyle[preprint,aps]{revtex}

\tightenlines

\begin{document}
\title{Quantum Computation with Coherent States, Linear Interactions and 
Superposed Resources.}
\author{T.~C.~Ralph~$^1$,  W.~J.~Munro~$^2$ and G.~J.~Milburn~$^{1,3}$\\
$^1$Centre for Quantum Computer Technology,
University of Queensland, Australia\\
$^2$Hewlett Packard Laboratories,
Filton Road, Stoke Gifford,\\
Bristol BS34 8QZ, U.K\\
$^3$~Institute for Quantum Information,
California Institute of Technology\\
MC 107-81, Pasadena, CA 91125-8100.
}
\date{\today}
\maketitle

\begin{abstract}

We show that quantum computation circuits using coherent states as the 
logical qubits can be constructed from very simple 
linear networks, conditional measurements and coherent superposition 
resource states.

\end{abstract}

\vspace{10 mm}

Quantum optics has proved a fertile field for experimental tests
of quantum information science, from experimental verification
of Bell inequality violations \cite{Bell1,Bell2}
to quantum teleportation \cite{teleport1}.
However, quantum optics was not thought to provide a practical 
path to efficient and scalable
quantum computation, and most current efforts to achieve this
have focussed on solid state implementations.
This orthodoxy was challenged recently when
Knill et al.\cite{KLM} showed that, given single photon sources
and single photon detectors, linear optics alone would suffice
to implement efficient quantum computation. While this result is
surprising, the complexity of the optical networks required is
daunting.

In this letter we propose an efficient scheme which is elegant in its
simplicity. By encoding the quantum information in multi-photon
coherent states, rather than single photon states, simple optical
manipulations acquire unexpected power. The required resource, which
may be produced non-deterministically, is a superposition of
the vacuum and a coherent state. Given this, the scheme is deterministic and
requires only simple linear optics and photon counting.
Qubit readout uses homodyne detection which can be highly efficient.

The idea of encoding quantum information on continuous variables 
of multi-photon fields has emerged recently \cite{teleport2} 
and a number of schemes 
have been proposed for realizing quantum computation in this way 
\cite{llo,sand,kim01}. One drawback of these proposals is that 
``hard'', non-linear interactions are required ``in-line'' of the 
computation. These would be very 
difficult to implement in practice. In contrast this proposal 
requires only ``easy', linear in-line interactions. The hard 
interactions are only required for ``off-line'' production of resource 
states. A related proposal is that of Gottesman et al \cite{pre} in 
which superpositions of squeezed states are used to encode the qubits. 
There the hard interactions are only used for the initial state preparation.
However, quadratic, squeezing type interactions, are required in-line 
along with linear interactions.

The output of a single mode, stabilised laser can be described by a
coherent state, $|\alpha \rangle$ where $\alpha$ is a complex number which
determines the average field amplitude.
Coherent states are defined by unitary transformation of the vacuum
\cite{WallsMilb94}, $|\alpha\rangle=D(\alpha)|0\rangle$, where
$D(\alpha)$ is the displacement operator.
Let us consider an encoding of logical qubits in coherent states
with ``binary pulse code modulation'',
\begin{eqnarray}
|0\rangle_L & = & |0\rangle\\
|1\rangle_L & = & |\alpha\rangle
\end{eqnarray}
where we take $\alpha$ to be real.
The advantage of using such states is that detection is relatively easy,
requiring only efficient
homodyne detection.

These qubits are not exactly orthogonal, but the approximation of
orthogonality is good for $\alpha$ even moderately large,
\begin{equation}
\langle \alpha |0\rangle =e^{-\alpha^2/2}
\label{orthog}
\end{equation}
We will assume for most of this paper that $\alpha >> 1$.

In single photon optics two qubit gates, in which the state of one
photon controls the state of the other, represent a formidable
challenge. Surprisingly, for our coherent state encoding, a
non-trivial two qubit gate can be implemented using only a single
beamsplitter. Consider the beamsplitter
interaction given by the unitary transformation
\begin{equation}
U_{BS}=\exp[i\theta (a b^{\dagger}+a^{\dagger} b)]
\end{equation}
where $a$ and $b$ are the annihilation operators corresponding to two
coherent state qubits $|\gamma \rangle_{a}$ and $|\beta
\rangle_{b}$, with $\gamma$ and $\beta$ taking values of $\alpha$ or
$0$. It is well known that the output state
produced by such an interaction is
\begin{eqnarray}
U_{BS} |\gamma \rangle_{a} |\beta \rangle_{b}=|\cos \theta \gamma+i
\sin \theta \beta \rangle_{a} |\cos \theta \beta+
i \sin \theta \gamma \rangle_{b}
 \label{Ho}
\end{eqnarray}
where $\cos^{2} \theta$ ($\sin^{2} \theta$) is the reflectivity
(transmissivity) of the beamsplitter.
Now consider the overlap between the output and input states. Using
the relationship \cite{WallsMilb94} $\langle \tau|\alpha \rangle =
\exp[-1/2(|\tau|^{2}+|\alpha|^{2})+\tau^{*} \alpha]$ we find
\begin{eqnarray}
\langle \gamma |_{a}\langle \beta |_{b} |\cos \theta \gamma+i
\sin \theta \beta \rangle_{a} |\cos \theta \beta+i \sin \theta \gamma
\rangle_{b} = \exp[-(\gamma^{2}+\beta^{2})(1-\cos \theta)+2 i \sin
\theta \gamma \beta]
\label{state}
\end {eqnarray}
Now suppose that $\theta$ is sufficiently small such that
$\theta^{2} \alpha^{2}<<1$ but that $\alpha$ is sufficiently large
that $\theta \alpha^{2}$ is of order one. Physically this corresponds
to an almost perfectly reflecting beamsplitter. Eq. \ref{state} then
approximately becomes
\begin{eqnarray}
\langle \gamma |_{a}\langle \beta |_{b} |\cos \theta \gamma+i
\sin \theta \beta \rangle_{a} |\cos \theta \beta+i \sin \theta \gamma
\rangle_{b} \approx \exp[2 i \theta \gamma \beta]
\label{astate}
\end {eqnarray}
Eq.\ref{astate} shows that the only difference between the input
and output states of the beamsplitter is a phase shift proportional
to the amplitudes of the input qubits, that is:
\begin{eqnarray}
U_{BS} |\gamma \rangle_{a} |\beta \rangle_{b}
& \approx & \exp[2 i \theta \gamma \beta] |\gamma \rangle_{a}
 |\beta \rangle_{b}
 \label{H3}
\end{eqnarray}

If conditions are such that Eq.\ref{H3} is a good approximation and we further
require that $\theta \alpha^{2}=\pi/2$ then this transformation produces
a controlled sign shift gate. That is if either or both of the qubits
are in the logical zero state ($\gamma=0$ and/or $\beta=0$)
the transformation produces no
effect on the state. However if both modes are initially in the logical one
state (i.e $\gamma=\beta=\alpha$) then a sign change is produced.
Such a gate is a non-trivial two qubit gate.

For universal computation we require, in addition to the two qubit gate above,
the ability to do arbitrary rotations that are diagonal in the computational
basis, bit-flip operations, plus the Hadamard gate \cite{NielChu}.
The Hadamard gate cannot be implemented unitarily with linear
optics. However, we will show shortly that, provided the necessary
quantum resource is possessed, it can be implemented using only
linear optics and conditional measurements.

First let us consider some single qubit transformations that can be
achieved with just linear optics. A {\em bit flip} gate flips the state of the
system from a
logical zero to a logical one, or vice versa and is equivalent to
the pauli $\sigma_x\equiv X$ matrix,
in the computational basis. The bit flip
transformation operator, $X$, is equivalent to a displacement of
$-\alpha$ followed by a $\pi$ phase shift of the coherent amplitude:
\begin{equation}
    X=U(\pi) D(-\alpha)
\end{equation}
where $U(\pi)=exp[i \pi a^{\dagger}a]$ is physically just a
half-wavelength delay, whilst a displacement can be implemented by
mixing a very strong coherent field with the qubit on a highly
reflective beamsplitter \cite{teleport2}.

The {\em phase rotation} gate produces a rotation
that is diagonal in the computational basis,
$ R_{\phi}(\mu|0 \rangle_L+\nu|1 \rangle_L)=\mu|0
    \rangle_L+e^{i \phi}\nu|1 \rangle_L$.
It can be implemented, to a good approximation, by imposing a
small phase shift on the qubit. Using arguments similar to
those leading to Eq.\ref{H3} we find
\begin{eqnarray}
    U(\epsilon)|\alpha \rangle & = & e^{i \epsilon a^{\dagger}a}|\alpha
\rangle \nonumber\\
& \approx & e^{i \epsilon \alpha^{2}}|\alpha
\rangle  =  R_{\phi}|\alpha \rangle
\label{phase}
\end{eqnarray}
with $\phi=\epsilon\alpha^2$. We have assumed $\epsilon$ scales as
$1/\alpha^2$.

In addition to these gates, we require a Hadamard gate in order to achieve an
arbitrary qubit rotation. The Hadamard gate, ${\cal H}$,
induces the following transformations on the logical states:
\begin{eqnarray}
    {\cal H}|0 \rangle_{L} & = & |0\rangle_L+|1\rangle_L=|0\rangle+|\alpha
    \rangle \nonumber\\
{\cal H}|1 \rangle_{L} & = &
|0\rangle_L-|1\rangle_L=|0\rangle-|\alpha  \rangle
    \label{cat}
    \end{eqnarray}
The outputs are a superposition of two widely
separated coherent states, commonly known as ``cat'' states.
Such states are highly non classical
and for unitary generation require a Kerr nonlinearity
for which the Hamiltonian is proportional to $(a^\dagger a)^2$. Such
interactions are typically very weak and
do not have sufficient strength to produce the required superposition
states. However we are not restricted to unitary transformations. A
number of schemes have been suggested which can produce parity cat states
non-deterministically \cite{song,welsch} and some experimental
progress has been made in their production \cite{Monro,Haroche,kimble}.
In all these schemes it is
necessary to distinguish between  a photon (or phonon) number
of $n$ and $n\pm 1$. If cat states could be used as a resource to
deterministically implement the Hadamard gate then these types of schemes
would
be sufficient for our purposes. We will now show this is true.

A Hadamard gate can be implemented using the two
qubit BS gate discussed in the previous section with one of the inputs
being the arbitrary state we wish to transform and the second input
being a known cat state. One of the outputs of the gate is measured
in the ``cat basis'' (see below)
and, depending on the result, a bit flip
operation may be required. This is a specific example of quantum gate
implementation via measurement. A general discussion of such
techniques can be found in Reference \cite{NielChu}.

A possible arrangement is shown in Fig.1.
Suppose the state we wish to transform, in the arbitrary state
$\mu|0 \rangle+\nu|\alpha \rangle$, is inserted into port 1 of the
BS gate whilst a resource cat state
$1/\sqrt{2}(|0\rangle+|\alpha \rangle)$ is inserted into port 2. The
output state of the gate is
\begin{equation}
   {{\mu}\over{\sqrt{2}}}(|0 \rangle_{1} |0 \rangle_{2}+|0 \rangle_{1}|\alpha
   \rangle_{2})+{{\nu}\over{\sqrt{2}}}(|\alpha \rangle_{1} |0
   \rangle_{2}-|\alpha \rangle_{1}
   |\alpha \rangle_{2})
   \end{equation}
Now suppose we make a measurement on output port 1 which returns a
dichotomic result telling us whether we have the same cat state as we
inserted or the (near) orthogonal state $1/\sqrt{2}(|0\rangle-|\alpha
\rangle)$.
If the result is the same cat state then the state of output port
2 is projected into
\begin{equation}
    {{1}\over{2}}(\mu+\nu)|0
    \rangle+{{1}\over{2}}(\mu-\nu)|\alpha \rangle
    \label{H}
\end{equation}
This is the required Hadamard transformation. On the other hand if the
opposite cat is measured at the output as was inserted then the projected
output state is
\begin{equation}
    {{1}\over{2}}(\mu-\nu)|0
    \rangle+{{1}\over{2}}(\mu+\nu)|\alpha \rangle
    \label{H-}
\end{equation}
But the state of Eq.\ref{H-} only differs from that of Eq.\ref{H} by a
bit flip operation. Thus the final step of the gate is to
implement (if
necessary) a bit flip on the output port.

A cat basis measurement
can be implemented in the following way. First we displace by
$-\alpha/2$. This transforms our ``0'', ``$\alpha$'' superposition into
``$\alpha/2$'', ``$-\alpha/2$'' superposition:
\begin{equation}
D(-\alpha/2)1/\sqrt{2}(|0\rangle \pm|\alpha
\rangle)=1/\sqrt{2}(|-\alpha/2\rangle \pm|\alpha/2 \rangle)
\end{equation}
These new states are parity eigenstates. Thus if photon number is
measured then an even result indicates detection of the state
$1/\sqrt{2}(|\alpha/2\rangle +|-\alpha/2 \rangle)$ and therefore
$1/\sqrt{2}(|0\rangle +|\alpha \rangle)$ whilst similarly an odd
result indicates detection of $1/\sqrt{2}(|0\rangle -|\alpha
\rangle)$ as can be confirmed by direct calculation. The cats could
also be distinguished by homodyne detection looking at the imaginary
quadrature \cite{next}. This latter 
technique would give inconclusive results some of the
time but may be useful for initial experimental demonstrations.

The control not gate (CNOT) is ubiquitous in quantum processing
tasks. It is also the simplest two-qubit gate whose operation can
easily be experimentally verified in the computational basis. A CNOT gate
will flip the state of one of the input qubits,
the ``target'', only if the other qubit, the ``control'', is in the
logical one state. If the control is in the logical zero state the
target is unchanged. A CNOT gate can be implemented as shown in Fig.2 by
first applying a Hadamard gate to the target state followed by the BS
gate applied to the control and target. Finally another Hadamard
gate is applied to the target.
For arbitrary control and target input
qubits we find:
\begin{equation}
   {\cal H}_{t} U_{BS} {\cal H}_{t} (\mu |0 \rangle+\nu |\alpha
   \rangle)_{c}(\gamma |0 \rangle+\tau |\alpha
   \rangle)_{t}=\mu \gamma |0 \rangle |0 \rangle+\mu \tau |0 \rangle
  |\alpha \rangle+
   \nu \tau |\alpha \rangle |0 \rangle+\nu \gamma |\alpha \rangle |\alpha
\rangle
\label{ccnot}
\end{equation}
which displays CNOT logic. The result of Eq.\ref{ccnot} assumes $\alpha>>1$.
To evaluate just how
large $\alpha$ needs to be we use the exact expression for the BS
gate, as given in Eq.\ref{Ho}, to calculate the output state of the CNOT.
We assume ideal bit flip operations and cat state preparation. Our
figure of merit is the average fidelity between the exact output and the
ideal output, as given by Eq.\ref{ccnot}.

The results are shown in Fig.3. In Fig.3(a) the average fidelity is
plotted as a function of $\alpha$. Fidelities of $0.9$ and above
require $\alpha > 10$. Such signal sizes, although commonplace in the
computational basis would be challenging to produce and
control in the superposition basis and the required technology is
probably some years away. On the other hand in Fig.3(b) a
renormalised average fidelity is plotted. This is obtained by
normalising the fidelity of getting the correct output state against
the sum of the fidelities for all the possible output states in the
computational subensemble. If there was no movement of states out of
this subensemble one would expect the two plots to be identical. The
fact that the renormalised fidelities remain high for much lower
values of $\alpha$ shows that qubit
leakage is the major reason for
the decreasing fidelities at moderate levels of $\alpha$ in Fig.3(a).
This in turn suggests that experimental demonstrations, albeit with
low efficiency, may be possible for $\alpha$'s as small as $3$.

The major sources of error in our scheme are expected to be, in order of
increasing significance:
(i) errors due to non orthogonal code states,
(ii) errors due to failure of the two qubit gate condition
($\theta^{2} \alpha^2<<1$),
(iii) erroneous identification of the input cat resource,
(iv) photon loss, and
(v) errors due to random optical phase shifts.
The first source of error becomes negligible for $\alpha>3$ (see
Eq.\ref{orthog}). Figure 3(a) shows that the second source of errors
is small for $\alpha >
20$. The third source is equivalent to a small rotation error in the code
space;
the fourth source causes a collapse to the one logical state,
while the final source is a phase error. It can be shown \cite{next} that
good quantum
error correction codes are available to correct
these  errors and further that error correction can be
implemented in a fault tolerant fashion.

In this letter we have presented a quantum computation scheme based
on encoding qubits as vacuum and coherent states, and their superposition.
The optical networks required are simple and compatible
with current optical communication networks. As well as the long term
goal of quantum computation, applications in quantum communication
protocols seem likely. Although the coherent amplitudes needed for scalable
computation are quite large our results indicate that experimental
demonstrations with modest amplitudes should be possible.

\noindent{\bf Acknowledgements.}
GJM acknowledges the support of the Institute for Quantum Information,
California Institute of Technology where this work was initiated. WJM
acknowledges funding in part by the European project
EQUIP(IST-1999-11053). The
Australian Research Council Special Research Centre for Quantum Computer
Technology supported this work.

\begin{figure}

\caption{Schematic of Hadamard gate based on the two qubit beamsplitter
gate (BS) and conditional implementation of the bit flip gate (F). If
an even number of photons is counted then the output is in the desired
state. If an odd number of photons is counted a bit flip operation is
needed to place the output in the correct state. }

\caption{Schematic of CNOT gate based on two qubit beamsplitter
gates (BS) and conditional implementation of bit flip gates (F).}

\caption{Performance of the CNOT gate as a function of the magnitude
of $\alpha$. In (a) the average fidelity is plotted against $\alpha$.
The high fidelities at very low values of $\alpha$ are an artefact of
the non-orthogonality of small $\alpha$ states. In (b) the fidelities
are renormalised against the total fidelity for a result within the
computational subensemble of states.}

\end{figure}

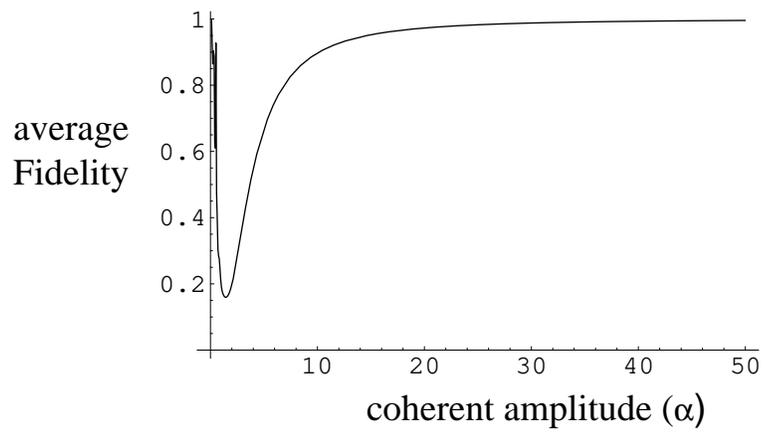
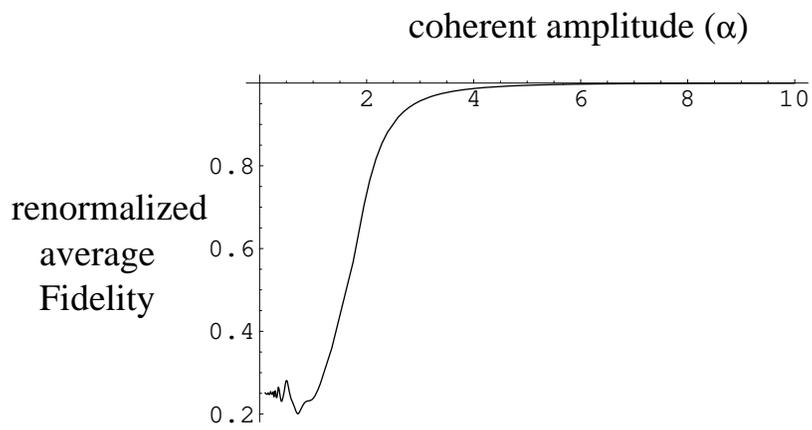
\end{document}